\documentstyle[12pt]{article}
\def\dfrac{\displaystyle\frac}
\def\i{\imath}
\def\a{\alpha}
\def\sn{{\rm sign}}
\def\bp{{\bf p}}
\def\d{\partial}

\def\noi{\noindent}
\def\ve{\varepsilon}
\def\ce{{\cal E}}

\def\br{\bar{\rho}}

\newcommand{\Eq}[1]{Eq.(\ref{#1})}

\newcommand{\refc}[1]{Ref.~\cite{#1}}

\newcommand{\bea}{\begin{eqnarray}}
\newcommand{\eea}{\end{eqnarray}}
\newcommand{\be}{\begin{equation}}
\newcommand{\ee}{\end{equation}}
\newcommand{\bc}{\begin{center}}
\newcommand{\ec}{\end{center}}
\newcommand{\ba}{\begin{array}}
\newcommand{\ea}{\end{array}}

\newcommand{\cL}{{\cal L}}
\newcommand{\Leff}{\cL^{\rm eff}(B,\mu)}
\newcommand{\Lb}{\cL^{\rm eff}(B)}
\newcommand{\Lm}{\tilde\cL^{\rm eff}(B,\mu)}

\newcommand{\ijmp}[3]{{\it Int. J. Mod. Phys. } {{\bf #1} {(#2)} {#3}}}

\newcommand{\np}[3]{{\it  Nucl. Phys. }{{\bf #1} {(#2)} {#3}}}

\newcommand{\prd}[3]{{\it  Phys. Rev. D} {{\bf #1} {(#2)} {#3}}}
\newcommand{\prl}[3]{ {\it Phys. Rev. Lett.}{{ \bf #1} {(#2)} {#3}}}
\newcommand{\pl}[3]{{\it  Phys. Lett. }{{\bf #1} {(#2)} {#3}}}

\newcommand{\sovjnp}[3]{{\it Sov. J. Nucl. Phys. }{{\bf #1} {(#2)} {#3}}}

\normalsize
\begin{document}

\large
\thispagestyle{empty}
\begin{flushright}                              FIAN/TD/96-22\\
                                                hep-th/9701100\\
                                                November 1996

\vspace{2cm}
\end{flushright}
\bc
\normalsize


{\LARGE\bf Induced Magnetic Field in a Finite Fermion Density
Maxwell QED$_{2+1}$}

\vspace{3ex}

{\Large Vadim Zeitlin$^{\dagger}$}

{\large Department of Theoretical Physics, P.~N.~Lebedev Physical
Institute,

  Leninsky prospect 53, 117924 Moscow, Russia}
\vspace{5ex}

\ec

\centerline{{\Large\bf Abstract}}

\normalsize
\begin{quote}
We are studying finite fermion density states in Maxwell QED$_{2+1}$
with external magnetic field. It is shown that at any
fermion density the energy of some magnetized states
 may be less than that of the state with the same density, but no magnetic
field.  Magnetized states are described by the effective
Maxwell-Chern-Simons QED$_{2+1}$ Lagrangian with gauge field mass proportional
to the number of filled Landau levels.
\end{quote}

\vfill
\noindent
$^\dagger$ E-mail address: zeitlin@lpi.ac.ru
\newpage


\setcounter{page}{1}

\section{Introduction}
For the last 15 years the study of the field theories in 3 space-time
dimensions is booming, with activity spanning from 3-dimensional  theories {\it
per se}, including anyon physics, to 3-dimensional euclidian models that
arise as high-temperature limit of
conventional (3+1)-dimensional theories, and to applications to planar condensed
matter systems (quantum and fractional Hall effects, high-temperature
superconductivity). In this letter we shall address ourselves to the problem of
spontaneous magnetization in finite fermion density  (2+1)-dimensional quantum
electrodynamics (QED$_{2+1}$).

In (2+1)-dimensional gauge models one may add to the usual bare Lagrangians
the so-called Chern-Simons term, which in the case of QED$_{2+1}$  is
proportional to $\ve_{\mu\nu\alpha}A^\mu\d^\nu A^\alpha$ \cite{DJT},

\medskip
        \be
        \cL = -\frac14 F_{\mu\nu}F^{\mu\nu} +
        \frac{\theta}4 \ve_{\mu\nu\alpha}F^{\mu\nu}A^\alpha +
        \bar{\psi}
        (\imath {\partial  \kern-0.5em/} + e {A\kern-0.5em/}   -m)\psi~~~.
        \label{mcs}
        \ee

\noi
The Chern-Simons term dramatically changes the theory making the gauge field
massive with the following propagator

        \be
        D_{\mu\nu} =\frac1{p^2 - \theta^2 +\i\ve}
        \left(
                \left( g_{\mu\nu} - \frac{p_\mu p_\nu}{p^2} \right)
                + \i \theta \ve_{\mu\nu\alpha}p^\alpha  \right)
                        +\alpha \frac{p_\mu p_\nu}{p^4}~~~.
        \label{dmunu0}
        \ee

\noi
In QED$_{2+1}$ with massive 2-component fermions  the
Chern-Simons term is generated dynamically in one-loop effective action even if
it is not present in the bare Lagrangian.

With the Lagrangian (1) the modified Maxwell equation becomes

        \be
        \d_\mu F^{\mu\nu} + \dfrac{\theta}2 \ve^{\nu\alpha\beta}F_{\alpha\beta}
        = j^\nu~~~,
        \label{mcseq}
        \ee

\noi
thus in the Maxwell-Chern-Simons (MCS) QED$_{2+1}$
electric and magnetic components are mixed up:
static charges produce magnetic field as well as electric one and
the electric charge $j^0_{cs} =  \theta B$ is associated with the uniform
magnetic field $B$. On the other hand, in an
external magnetic field
a term corresponding to induced charge may arise
in the effective action $\Leff = -\i {\rm Tr} \log
(\imath {\partial  \kern-0.5em/} + e {A\kern-0.5em/}   -m)$.
For zero
chemical potential in QED$_{2+1}$ the charge is  $j^0_f = -
\frac{e^2B}{4\pi}$ \cite{NS,R}, while for finite chemical potential it is $j^0_f
= - \frac{e^2B}{4\pi} (2L+1)$, $L$ is the number of
filled Landau levels [4-7].  Therefore, in MCS
QED$_{2+1}$ with an external magnetic field the electric neutrality condition
is $j^0_f + j^0_{cs} = 0$.
Several years ago Hosotani   has shown that in MCS QED$_{2+1}$ the above neutral
configuration with a uniform magnetic field within the two-loop approximation
may have energy lesser than the naive vacuum \cite{H93,H95}.  In this case
magnetic field arises via Goldstone mechanism since the condition $j^0_f +
j^0_{cs} = 0$ entails the mass of the gauge field to be vanishing. That
approach was extended in Refs. [10-12], in particular, to finite-temperature
case.

        In this paper we shall investigate  the possibility of the
magnetization within the finite fermion density  QED$_{2+1}$ without
any bare Chern-Simons term, i.e. within the Maxwell QED$_{2+1}$, which was
discussed in brief in the recent talk \cite{Z96}.  Our approach is based on the
observation that the equation for the fermion density $\rho(B,\mu) = {\rm
const}.$ has an infinite number of solutions and one may minimize the energy of
a fixed density configuration by varying the magnetic field. The paper is
organized as follows.  In Sec. 2 the fermion density as a function of chemical
potential and external magnetic field is obtained basing on the spectral
properties.
In Sec. 3 we are
minimizing the energy of the fixed fermion density states by varying the
magnetic filed to show that the magnetization may be energetically preferable.
In Sec. 4 the effective theory for  the finite
fermion density QED$_{2+1}$ with magnetic field is considered and it is shown
that the magnetized states are described by the Maxwell-Chern-Simons
Lagrangian.  Concluding remarks are presented in Sec.  5.

\section{Fermion density in QED$_{2+1}$ with a uniform magnetic field}

Below we shall consider the Maxwell  QED$_{2+1}$ with two-component spinors
($\gamma$-matrices are the Pauli matrices, $\gamma^0 = \sigma_3, \gamma^{1,2} =
\i\sigma_{1,2}$),

        \be
        \cL = -\frac14 F_{\mu\nu}F^{\mu\nu} +
        \bar{\psi}(\imath {\partial
        \kern-0.5em/} + e {A\kern-0.5em/}   -m)\psi~~~,
        \label{maxwell}
        \ee

\noi
with a uniform magnetic field $B$, $F_{12}=-F_{21} =- B$
present in the system. In this case, as we have mentioned, the induced charge
arises, $j^0_f = -  \frac{e^2B}{4\pi}$ \cite{NS,R}. This charge (fermion
density) arises due to the asymmetry of the fermion spectrum of
QED$_{2+1}$ in an external magnetic field \cite{NS}:
the fermion density is equal to {\em degeneracy} $\times$ {\em fermion number}
and the latter is defined by the fermion spectrum,

        \be
        N = - {1 \over 2} \sum_{p_0^{(n)}} {\rm sign} (p^{(n)}_0)~~~.
        \label{b_number}
        \ee

The fermion spectrum (Landau levels) in the presence of a uniform magnetic field
is asymmetric,

        \be
        p_0^{(0)} = -m~\sn (eB), \quad
        p_0^{(\pm n)} = \pm  \sqrt{m^2 + 2|eB|n}, \quad n= 1,2, \dots ,
        \label{spectrum}
        \ee

\noi
providing $N=\frac12$, and the degeneracy of every  energy level is
$\frac{|eB|}{2\pi}$.
Thus, as soon as the external magnetic field emerges in the Maxwell QED$_{2+1}$
it gives rise the fermion density.

To gain another parameter to vary
the fermion density independently we introduce chemical potential $\mu$ by
adding the term (-$\mu\psi^\dagger\psi$) to the Lagrangian (\ref{maxwell})
(and modifying the $\i\ve$--prescription in the fermion Green function).  The
fermion density in QED$_{2+1}$ with an external magnetic field and finite
chemical potential was calculated using different approaches
[5-7], but the most straightforward procedure is based on the
spectral properties of the theory:  equation (\ref{b_number}) may be
extended to the case of finite chemical potential at $T=0$ as \cite{N85}

        \be
        N = - {1 \over 2} \sum_{p_0^{(n)}} {\rm sign} (p^{(n)}_0) +
        \sum_{p_0^{(n)}}
        \left(
        \theta(p^{(n)}_0)\theta(\mu - p^{(n)}_0) -
        \theta(- p^{(n)}_0)\theta(p^{(n)}_0 - \mu)
                        \right)\quad.
        \label{mu_number}
        \ee

The latter combined with \Eq{spectrum} yields the following:

        \bea
        \rho (B,\mu) = \frac{eB}{4\pi} +
        \left\{
        \begin{array}{cl}
        {}~~\dfrac{|eB|}{2\pi}
        \left(\left[
        \dfrac{\mu^2 -m^2}{2|eB|} \right]
        + \theta(-eB)                   \right),\quad     & \mu>\/m;
        \\
        &\\
        0,                   \quad             & |\mu|<m;\\
        &\\
        {}-
        \dfrac{|eB|}{2\pi}
        \left(
        \left[
        \dfrac{\mu^2 - m^2}{2|eB|}      \right]
        + \theta(eB)        \right) ,\quad &\mu<- m,
        \end{array}\right.
        \label{rho}
        \eea

\noi
where $[ \dots ]$ denotes the integral part,
 $\left[ \frac{\mu^2 -m^2}{2|eB|} \right]$ or ($\left[ \frac{\mu^2
-m^2}{2|eB|} \right] +1$) is the number of filled Landau levels
(at zero temperature in terms of $\mu$ and $B$ one can describe
completely filled Landau levels only).
Due to the spectral asymmetry the density is asymmetric, too. The discreteness
of the spectrum makes the density at zero temperature (discontinuous) step-like
function, Fig. 1 (in $(\rho,B)$-plane it is saw-tooth like). In the $B
\rightarrow 0$ limit the density is

\be
\rho(\mu) = \frac1{4\pi}(\mu^2-m^2)\theta(\mu^2-m^2)\sn (\mu)~~~.
\label{rhomu}
\ee

\begin{figure}
\unitlength=1.00mm
\linethickness{0.4pt}
\begin{picture}(143.00,111.83)
\put(55.00,70.11) {\makebox(0,0)[cc]{{\tiny $\bullet$}}}
\put(45.00,70.11) {\makebox(0,0)[cc]{{\tiny $\bullet$}}}
\put(35.00,70.11) {\makebox(0,0)[cc]{{\tiny $\bullet$}}}
\put(105.00,70.11){\makebox(0,0)[cc]{{\tiny $\bullet$}}}
\put(115.00,70.11){\makebox(0,0)[cc]{{\tiny $\bullet$}}}
\put(125.00,70.11){\makebox(0,0)[cc]{{\tiny $\bullet$}}}
\put(135.00,70.11){\makebox(0,0)[cc]{{\tiny $\bullet$}}}
\put(80.00,36.99){\vector(0,1){74.84}}
\put(27.00,70.11){\vector(1,0){116.00}}
\put(140.00,64.95){\makebox(0,0)[cc]{$\mu$}}
\put(115.00,108.82){\makebox(0,0)[rc]{$\rho(B=const,~\mu)$}}
\put(108.00,67.10){\makebox(0,0)[cc]{{\scriptsize $p^{(1)}_0$}}}
\put(118.00,67.10){\makebox(0,0)[cc]{{\scriptsize $p^{(2)}_0$}}}
\put(128.00,67.10){\makebox(0,0)[cc]{{\scriptsize $p^{(3)}_0$}}}
\put(51.00,67.50){\makebox(0,0)[cc]{{\scriptsize $p^{(-1)}_0$}}}
\put(41.00,67.50){\makebox(0,0)[cc]{{\scriptsize $p^{(-2)}_0$}}}
\put(31.00,67.50){\makebox(0,0)[cc]{{\scriptsize $p^{(-3)}_0$}}}
\put(80.00,74.84) {\makebox(0,0)[cc]{{\tiny $\bullet$}}}
\put(80.00,85.16) {\makebox(0,0)[cc]{{\tiny $\bullet$}}}
\put(80.00,95.05) {\makebox(0,0)[cc]{{\tiny $\bullet$}}}
\put(80.00,55.05) {\makebox(0,0)[cc]{{\tiny $\bullet$}}}
\put(80.00,43.87) {\makebox(0,0)[cc]{{\tiny $\bullet$}}}
\put(86.00,95.05){\makebox(0,0)[cc]{$\frac{5eB}{4\pi}$}}
\put(86.00,85.16){\makebox(0,0)[cc]{$\frac{3eB}{4\pi}$}}
\put(87.00,55.05){\makebox(0,0)[cc]{$-\frac{3eB}{4\pi}$}}
\put(87.00,43.87){\makebox(0,0)[cc]{$-\frac{5eB}{4\pi}$}}
\put(65.00,70.11) {\makebox(0,0)[cc]{{\tiny $\bullet$}}}
\put(80.00,64.95) {\makebox(0,0)[cc]{{\tiny $\bullet$}}}
\put(95.00,70.11) {\makebox(0,0)[cc]{{\tiny $\circ$}}}
\put(61.00,67.50){\makebox(0,0)[cc]{{\scriptsize $p^{(0)}_0$}}}
\put(65.00,74.84){\line(1,0){40}}
\put(105.00,85.16){\line(1,0){10}}
\put(115.00,95.05){\line(1,0){10}}
\put(55.00,64.95){\line(1,0){10}}
\put(45.00,55.05){\line(1,0){10}}
\put(35.00,43.87){\line(1,0){10}}
\put(87.00,64.95){\makebox(0,0)[cc]{$-\frac{eB}{4\pi}$}}
\end{picture}

\vspace{-3cm}
\caption{Fermion density as a function of chemical potential, $B$=const.}

 \end{figure}
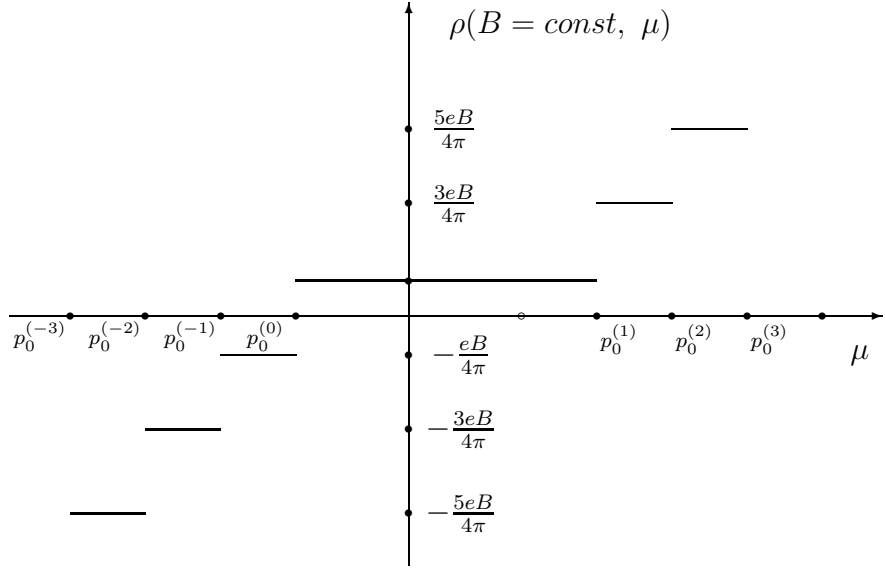

With the expression for the fermion density at hand one may easily obtain the
$\mu$-dependent part of the one-loop effective Lagrangian,
$\Lm = {\displaystyle \int_0^\mu}\rho(B,\mu')d\!\mu'$. The full effective
Lagrangian is $\Leff = \Lb + \Lm$, where $\Lb$ is QED$_{2+1}$ Heisenberg-Euler
effective Lagrangian calculated in \refc{R},

        \be
        \Lb = \dfrac1{8\pi^{3/2}} \int_0^\infty \dfrac{d\!s}{s^{5/2}}
        e^{-m^2s} (eBs \coth(eBs) -1)~~~.
        \label{lb}
        \ee

\section{Magnetization in the finite fermion density QED$_{2+1}$}

Our further consideration is based on the observation that at zero temperature
in QED$_{2+1}$ with $B \ne 0, |\mu| > |m|$ the equation $\rho(B,\mu) = {\rm
const.}$ has an infinite number of solutions, enumerated by the filled Landau
levels number \cite{Z95}. Indeed, when the (positive) chemical potential is less
than the (positive) lowest Landau level energy (we are suggesting $B, \mu >0$
below),
$0 < \mu < \sqrt{m^2 + 2eB}$, the fermion density is $\rho = \frac{eB_0}{4\pi}$ .
For
$\sqrt{m^2 + 2eB} < \mu < \sqrt{m^2 + 4eB}$ the density is $\rho =
\frac{3eB_1}{4\pi}$ etc,  i.e. we may consider the states of the same
density constituted  of  different numbers of filled Landau levels (let us
stress again that at $T=0$ considering of completely filled Landau levels
is only permitted). In $(\mu^2 - m^2,B)$--plane these solutions of the equation
$\rho(B,\mu) = {\rm const.}$ are the segments parallel to the $\mu$--axis, Fig.
2, and the magnetic field corresponding to the configuration with $L$ filled
Landau levels and fermion density $\rho$ is

        \be
        eB_L = \dfrac{4\pi\rho}{2L+1}, \quad L=0,1,2,\dots~~~,
        \label{ebl}
        \ee

\noi
while the chemical potential is not uniquely defined.

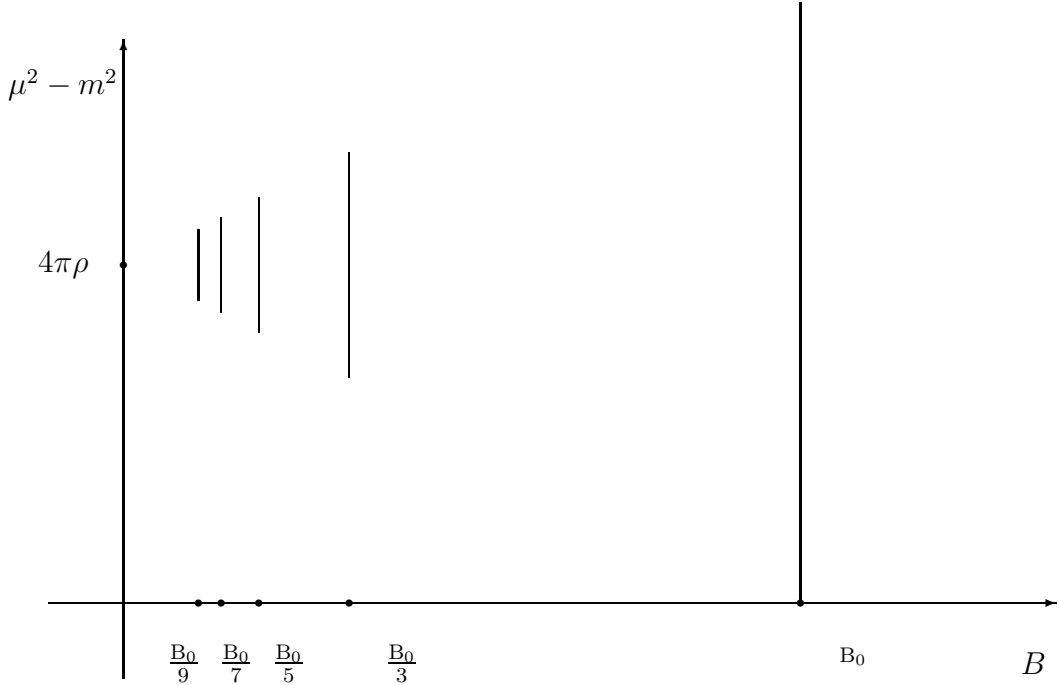
\begin{figure}
\unitlength=1.00mm
\linethickness{0.4pt}

\vspace{-2cm}
\begin{picture}(145.00,125.00)
\put(20.00,20.00){\vector(0,1){85.00}}
\put(10.00,30.00){\vector(1,0){134.00}}
\put(110.00,30.00){\line(0,1){80}}
\put(50.00,60.00){\line(0,1){30}}
\put(38.00,66.00){\line(0,1){18}}
\put(33.00,68.75){\line(0,1){12.50}}
\put(30.00,70.25){\line(0,1){9.50}}
\put(141.00,22.00){\makebox(0,0)[cc]{$B$}}
\put(12.00,99.00){\makebox(0,0)[cc]{$\mu^2 - m^2$}}
\put(20.00,75.00){\makebox(0,0)[cc]{{\tiny $\bullet$}}}
\put(12.00,75.00){\makebox(0,0)[cc]{$4\pi\rho$}}
\put(117.00,23.00){\makebox(0,0)[cc]{\scriptsize{${\rm B}_0$}}}
\put(57.00,22.00){\makebox(0,0)[cc]{$\frac{{\rm B}_0}{3}$}}
\put(42.00,22.00){\makebox(0,0)[cc]{$\frac{{\rm B}_0}{5}$}}
\put(35.00,22.00){\makebox(0,0)[cc]{$\frac{{\rm B}_0}{7}$}}
\put(28.00,22.00){\makebox(0,0)[cc]{$\frac{{\rm B}_0}{9}$}}
\put(110.00,30.00){\makebox(0,0)[cc]{{\tiny $\bullet$}}}
\put(50.00,30.00) {\makebox(0,0)[cc]{{\tiny $\bullet$}}}
\put(38.00,30.00) {\makebox(0,0)[cc]{{\tiny $\bullet$}}}
\put(33.00,30.00) {\makebox(0,0)[cc]{{\tiny $\bullet$}}}
\put(30.00,30.00) {\makebox(0,0)[cc]{{\tiny $\bullet$}}}
\end{picture}

\vspace{-2.cm}
\caption{Solutions of the equation $\rho(B,\mu)= \rho$
in $(B,\mu^2 - m^2)$--plane, ~~$eB_0 = 4\pi\rho$}
\end{figure}

Thus, after the density of the fermion gas is fixed we still have freedom to
choose a magnetic field. Below we shall investigate the possibility of
spontaneous magnetization in a finite fermion density QED$_{2+1}$ by minimizing
the energy of the configurations with equal density with respect to the
magnetic field (number of filled Landau levels).  If the energy of any
configuration is less than the energy of the configuration with the same
density, but no magnetic field, spontaneous magnetization takes place for that
value of the fermion density (and coupling constant).

The relevant calculations are straightforward. Making the Legandre
transformation one can obtain the energy density

        \be
        \ce(B,\rho) = \mu\rho - \Leff = \mu\rho +\frac{B^2}2 - \Lb - \Lm~~~.
        \ee

In the weak-field limit $eB/m^2 \ll 1$ one may evaluate the integral \Eq{lb}
and the energy density $\ce_L$ of
the configuration with specific $L$ is the following

        \be
        \ce_L(\rho) = \dfrac{B^2}2
        \left(  1 - \dfrac{e^2}{12\pi m}        \right)
        + \dfrac{eB}{2\pi} \sum_{n=1}^L \sqrt{m^2 + 2eBn}
        \label{en1}
        \ee

\noi
(the latter term in the right-hand side of \Eq{en1} is just the sum of the
energies of the filled Landau levels $\times$ degeneracy). $L$ is uniquely
defined by the values of $\rho$ and $B$ and the energy density may be rewritten
in terms of $\rho$ and $L$ as

        \be
        \ce_L (\rho) = \dfrac{8\pi^2\rho^2}{e^2(2L+1)^2}
        \left(  1 - \dfrac{e^2}{12\pi m}        \right)
        + \dfrac{2\rho}{2L+1}
        \sum_{n=1}^L \sqrt{m^2 + \dfrac{8\pi\rho n}{2L+1}}~~~.
        \label{en2}
        \ee

For the vanishing magnetic field the energy density  is the following:

        \be
        \ce(\rho)= \frac1{6\pi} (\mu^3 - m^3)\theta(\mu^2-m^2) =
        \frac1{6\pi}((m^2+4\pi\rho)^{3/2} - m^3)~~~
        \label{emu1}
        \ee

\noi
and magnetization condition  $\ce_L (\rho) <\ce (\rho)$ in terms of
dimensionless variables $\br=\rho/m^2$ and $\alpha = e^2/8\pi m$ is:

        \be
        \dfrac1{6\pi}\left( (1+4\pi\br)^{3/2} - 1 \right) >
        \dfrac{\pi\br^2}{\alpha(2L+1)^2} (1-{2\over 3}\alpha)
        + \dfrac{2\br}{2L+1}
        \sum_{n=1}^L \sqrt{1+\dfrac{8\pi\br n}{2L+1}}~~~.
        \label{mag1}
        \ee

   Let us start with the low density case,  $\br \ll 1$. Making expansion in
powers of $\br$ in the inequality (\ref{mag1}) one has

        \be
        \dfrac1{6\pi} (6\pi\br + {3\over 8} (4\pi\br)^2) >
        \dfrac{\pi\br^2}{\a (2L+1)^2} (1-{2\over 3}\a)
        + \dfrac{2\br}{2L+1}
        \sum_{n=1}^L
        \left(   1 + \dfrac{4\pi\br n}{2L+1}     \right)+{\cal O}(\br^3)~~~.
        \label{mag2}
        \ee

\noi
The latter inequality does have solutions if $\pi\br < \alpha(2L+1)$.

Thus we have shown that in the finite density QED$_{2+1}$ magnetization
may be energetically preferable, Fig. 3. The energy is minimum providing that
$\left[\frac{\pi\br}{\a}\right]$ or ($\left[\frac{\pi\br}{\a}\right] +1$)
Landau levels are filled. For instance, at $\rho < \frac{3\alpha}{4\pi}$
the induced magnetic field is $B=\frac{4\pi\rho}{e}$ (and for $\rho
\rightarrow 0$ the field vanishes). For $\frac{3\alpha}{4\pi} <\rho <
\frac{15\alpha}{8\pi}$  the field is $B=\frac{4\pi\rho}{3e}$, etc.

\begin{figure}
\unitlength=1.00mm
\linethickness{0.4pt}

\vspace{-9.5cm}
\hspace{1.cm}\begin{picture}(120.00,110.00)

\put(00.00,-80.00){\vector(0,1){100.00}}
\put(00.00,00.00){\vector(1,0){116.00}}

\put(83.77,15.38) {\makebox(0,0)[cc]{{\tiny $\bullet$}}}
\put(68.54,-39.29) {\makebox(0,0)[cc]{{\tiny $\bullet$}}}
\put(58.00,-63.63) {\makebox(0,0)[cc]{{\tiny $\bullet$}}}
\put(50.26,-74.46) {\makebox(0,0)[cc]{{\tiny $\bullet$}}}
\put(44.35,-78.73) {\makebox(0,0)[cc]{{\tiny $\bullet$}}}
\put(39.68,-79.65) {\makebox(0,0)[cc]{{\tiny $\bullet$}}}
\put(35.90,-78.81) {\makebox(0,0)[cc]{{\tiny $\bullet$}}}
\put(32.78,-77.04) {\makebox(0,0)[cc]{{\tiny $\bullet$}}}
\put(30.16,-74.81) {\makebox(0,0)[cc]{{\tiny $\bullet$}}}
\put(27.93,-72.36) {\makebox(0,0)[cc]{{\tiny $\bullet$}}}
\put(26.00,-69.86) {\makebox(0,0)[cc]{{\tiny $\bullet$}}}
\put(24.32,-67.38) {\makebox(0,0)[cc]{{\tiny $\bullet$}}}
\put(22.85,-64.97) {\makebox(0,0)[cc]{{\tiny $\bullet$}}}
\put(21.54,-62.66) {\makebox(0,0)[cc]{{\tiny $\bullet$}}}
\put(20.38,-60.45) {\makebox(0,0)[cc]{{\tiny $\bullet$}}}
\put(19.33,-58.35) {\makebox(0,0)[cc]{{\tiny $\bullet$}}}
\put(18.39,-56.37) {\makebox(0,0)[cc]{{\tiny $\bullet$}}}
\put(17.53,-54.49) {\makebox(0,0)[cc]{{\tiny $\bullet$}}}
\put(16.76,-52.72) {\makebox(0,0)[cc]{{\tiny $\bullet$}}}
\put(16.04,-51.04) {\makebox(0,0)[cc]{{\tiny $\bullet$}}}
\put(15.39,-49.46) {\makebox(0,0)[cc]{{\tiny $\bullet$}}}
\put(14.78,-47.96) {\makebox(0,0)[cc]{{\tiny $\bullet$}}}
\put(14.23,-46.55) {\makebox(0,0)[cc]{{\tiny $\bullet$}}}
\put(13.71,-45.21) {\makebox(0,0)[cc]{{\tiny $\bullet$}}}
\put(13.23,-43.94) {\makebox(0,0)[cc]{{\tiny $\bullet$}}}
\put(12.78,-42.73) {\makebox(0,0)[cc]{{\tiny $\bullet$}}}
\put(12.36,-41.56) {\makebox(0,0)[cc]{{\tiny $\bullet$}}}
\put(11.97,-40.50) {\makebox(0,0)[cc]{{\tiny $\bullet$}}}
\put(11.60,-39.47) {\makebox(0,0)[cc]{{\tiny $\bullet$}}}
\put(11.25,-38.48) {\makebox(0,0)[cc]{{\tiny $\bullet$}}}
\put(10.93,-37.55) {\makebox(0,0)[cc]{{\tiny $\bullet$}}}
\put(10.62,-36.65) {\makebox(0,0)[cc]{{\tiny $\bullet$}}}
\put(10.33,-35.80) {\makebox(0,0)[cc]{{\tiny $\bullet$}}}
\put(10.05,-34.98) {\makebox(0,0)[cc]{{\tiny $\bullet$}}}
\put(9.79,-34.20) {\makebox(0,0)[cc]{{\tiny $\bullet$}}}
\put(9.54,-33.45) {\makebox(0,0)[cc]{{\tiny $\bullet$}}}
\put(9.31,-32.73) {\makebox(0,0)[cc]{{\tiny $\bullet$}}}
\put(9.08,-32.04) {\makebox(0,0)[cc]{{\tiny $\bullet$}}}
\put(8.87,-31.38) {\makebox(0,0)[cc]{{\tiny $\bullet$}}}
\put(8.67,-30.75) {\makebox(0,0)[cc]{{\tiny $\bullet$}}}
\put(8.47,-30.14) {\makebox(0,0)[cc]{{\tiny $\bullet$}}}
\put(8.29,-29.56) {\makebox(0,0)[cc]{{\tiny $\bullet$}}}
\put(8.11,-28.99) {\makebox(0,0)[cc]{{\tiny $\bullet$}}}
\put(7.94,-28.45) {\makebox(0,0)[cc]{{\tiny $\bullet$}}}
\put(7.77,-27.93) {\makebox(0,0)[cc]{{\tiny $\bullet$}}}
\put(7.62,-27.42) {\makebox(0,0)[cc]{{\tiny $\bullet$}}}
\put(7.47,-26.93) {\makebox(0,0)[cc]{{\tiny $\bullet$}}}

  \put(55.85, -15.52) {\makebox(0,0)[cc]{{\tiny $*$}}}
  \put(45.70, -43.45) {\makebox(0,0)[cc]{{\tiny $*$}}}
  \put(38.67, -54.78) {\makebox(0,0)[cc]{{\tiny $*$}}}
  \put(33.51, -58.92) {\makebox(0,0)[cc]{{\tiny $*$}}}
  \put(29.57, -59.71) {\makebox(0,0)[cc]{{\tiny $*$}}}
  \put(26.46, -58.88) {\makebox(0,0)[cc]{{\tiny $*$}}}
  \put(23.94, -57.27) {\makebox(0,0)[cc]{{\tiny $*$}}}
  \put(21.85, -55.31) {\makebox(0,0)[cc]{{\tiny $*$}}}
  \put(20.11, -53.21) {\makebox(0,0)[cc]{{\tiny $*$}}}
  \put(18.62, -51.11) {\makebox(0,0)[cc]{{\tiny $*$}}}
  \put(17.33, -49.06) {\makebox(0,0)[cc]{{\tiny $*$}}}
  \put(16.21, -47.09) {\makebox(0,0)[cc]{{\tiny $*$}}}
  \put(15.23, -45.23) {\makebox(0,0)[cc]{{\tiny $*$}}}
  \put(14.36, -43.48) {\makebox(0,0)[cc]{{\tiny $*$}}}
  \put(13.59, -41.82) {\makebox(0,0)[cc]{{\tiny $*$}}}
  \put(12.89, -40.27) {\makebox(0,0)[cc]{{\tiny $*$}}}
  \put(12.26, -38.82) {\makebox(0,0)[cc]{{\tiny $*$}}}
  \put(11.69, -37.46) {\makebox(0,0)[cc]{{\tiny $*$}}}
  \put(11.17, -36.18) {\makebox(0,0)[cc]{{\tiny $*$}}}
  \put(10.69, -34.97) {\makebox(0,0)[cc]{{\tiny $*$}}}
  \put(10.26, -33.84) {\makebox(0,0)[cc]{{\tiny $*$}}}
  \put(9.86 , -32.78) {\makebox(0,0)[cc]{{\tiny $*$}}}
  \put(9.48 , -31.78) {\makebox(0,0)[cc]{{\tiny $*$}}}
  \put(9.14 , -30.83) {\makebox(0,0)[cc]{{\tiny $*$}}}
  \put(8.82 , -29.93) {\makebox(0,0)[cc]{{\tiny $*$}}}
  \put(8.52 , -29.09) {\makebox(0,0)[cc]{{\tiny $*$}}}
  \put(8.24 , -28.29) {\makebox(0,0)[cc]{{\tiny $*$}}}
  \put(7.98 , -27.53) {\makebox(0,0)[cc]{{\tiny $*$}}}
  \put(7.73 , -26.81) {\makebox(0,0)[cc]{{\tiny $*$}}}
  \put(7.50 , -26.12) {\makebox(0,0)[cc]{{\tiny $*$}}}
  \put(7.28 , -25.47) {\makebox(0,0)[cc]{{\tiny $*$}}}
  \put(7.08 , -24.85) {\makebox(0,0)[cc]{{\tiny $*$}}}
  \put(6.89 , -24.26) {\makebox(0,0)[cc]{{\tiny $*$}}}
  \put(6.70 , -23.69) {\makebox(0,0)[cc]{{\tiny $*$}}}
  \put(6.53 , -23.15) {\makebox(0,0)[cc]{{\tiny $*$}}}
  \put(6.36 , -22.63) {\makebox(0,0)[cc]{{\tiny $*$}}}
  \put(6.21 , -22.14) {\makebox(0,0)[cc]{{\tiny $*$}}}
  \put(6.06 , -21.67) {\makebox(0,0)[cc]{{\tiny $*$}}}
  \put(5.91 , -21.21) {\makebox(0,0)[cc]{{\tiny $*$}}}
  \put(5.78 , -20.78) {\makebox(0,0)[cc]{{\tiny $*$}}}
  \put(5.65 , -20.36) {\makebox(0,0)[cc]{{\tiny $*$}}}
  \put(5.52 , -19.96) {\makebox(0,0)[cc]{{\tiny $*$}}}
  \put(5.40 , -19.57) {\makebox(0,0)[cc]{{\tiny $*$}}}
  \put(5.29 , -19.20) {\makebox(0,0)[cc]{{\tiny $*$}}}
  \put(5.18 , -18.84) {\makebox(0,0)[cc]{{\tiny $*$}}}
  \put(5.08 , -18.49) {\makebox(0,0)[cc]{{\tiny $*$}}}

\put(110.00,-5.00){\makebox(0,0)[cc]{$eB$}}
\put(-5.00,15.00){\makebox(0,0)[cc]{$\Delta{\cal E}$}}

\end{picture}

\vspace{8.2cm}
\caption{Energy difference, $\Delta\ce = \ce_L(\rho) - \ce(\rho)$ for
$\br=0.003, \a = 0.001$ (dots) and $\br=0.002, \a = 0.00075$ (stars)
($L= 4, \dots,  50$, $eB_L=4\pi\rho/(2L +1)$).}


 \end{figure}
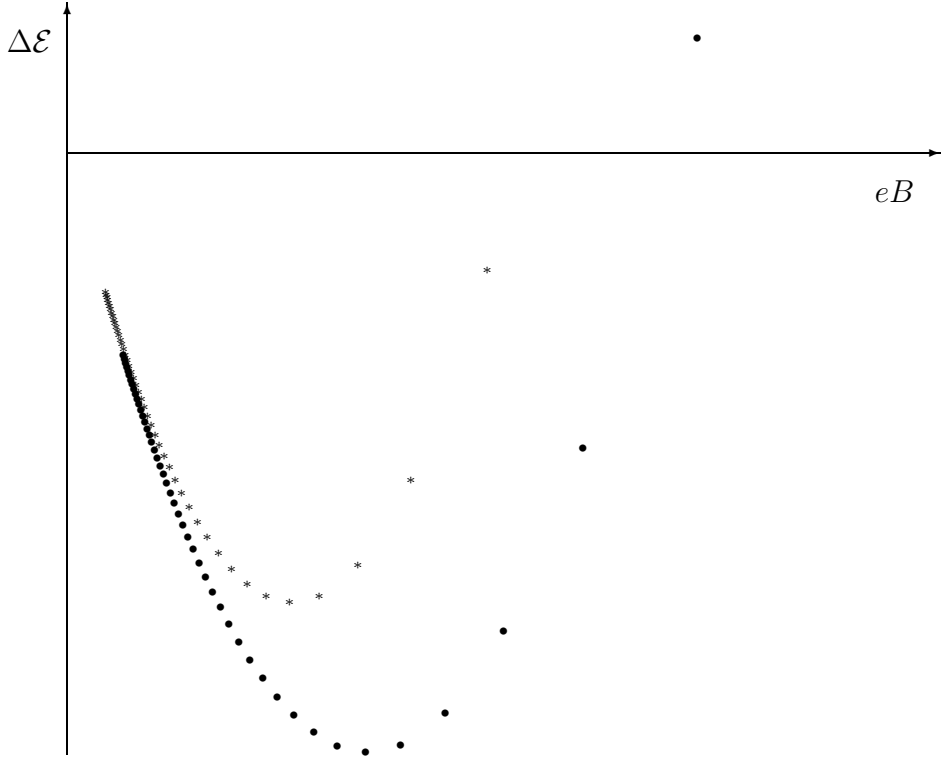

Since the fermion spectrum is asymmetric, the case $\mu
>0, B<0$ should be examined separately. The (positive) fermion density now
is $\rho = \frac{|eB|}{2\pi} (2L-1), L=1,2,\dots$ and the energy density is the
following:

        \be
        \ce_L = \dfrac{8\pi^2\rho^2}{e^2(2L+1)^2}
        \left(
        1- \frac{e^2}{12\pi m}                  \right)
        +\dfrac{2\rho}{2L+1}
        \sum_{n=0}^L \sqrt{m^2+\dfrac{8\pi\rho n}{2L+1}}
        \label{en-}
        \ee

\noi
and the magnetization condition $\ce_\mu > \ce_L$ is

        \be
        \br + \pi\br^2 > \dfrac{\pi\br^2}{\a(2L+1)^2}
        + \dfrac{2L+2}{2L+1} \br +\dfrac{4(L+1)(L+2)}{(2L+1)^2}\pi\br^2
        +{\cal O}(\br^3)~~~,
        \ee
\noi
which has no solutions, i.e. for $\mu > 0$ (finite {\em particle}
density) in the case of magnetization the field has a definite sign ($B>0$).
It is obvious that for $\mu < 0$ (finite {\em
antiparticle} density) the field would have the opposite sign ($B<0$).

However, magnetization  takes place not only at a low fermion density.
Supposing $L$ to be a big number one may use Euler-Maclaurin equation to
replace the sum in \Eq{en2}. The inequality
$\ce_L(\rho) < \ce(\rho)$ may be rewritten as

        \bea
        \frac1{6\pi}\left( (1+4\pi\br)^{3/2} - 1 \right)
        - \frac1{6\pi}\left( (1+\frac{8\pi\br L}{2L+1})^{3/2} - 1 \right)
        - \frac{\br}{2L+1}\left( (1+\frac{8\pi\br L}{2L+1})^{1/2} - 1 \right)
        >\nonumber\\
        {}\\
        \dfrac{\pi\br^2}{\alpha(2L+1)^2} (1-{2\over 3}\alpha)
        +{\cal O}\left(\frac{\br^2}{(2L+1)^2}\right)~~~~~~~~~~~~~~
        \nonumber\\
        \eea

\noi
and in the leading order
        \be
        \dfrac{\br}{2L+1} > \dfrac{\pi\br^2}{\a(2L+1)^2} +
        {\cal O}(\frac{\pi\br^2}{(2L+1)^2})~~~,
        \ee

\noi
i.e. the magnetization condition is $\pi\br< \a (2L+1)$. Even at high density
provided $\pi\br< \a (2L+1)$ both $\frac{\br}{2L+1}$ and  the magnetic field
are defined by the coupling constant and small, thus the weak field and
Euler-Maclaurin expansions are justified.

\section{Effective theory for the magnetized Maxwell QED$_{2+1}$}

In this section we shall consider one-loop effective theory for the
magnetized  finite density QED$_{2+1}$.  In QED$_{2+1}$ at $B,\mu\ne0$ it is
possible to gain insight into the low-momentum behavior using just general
properties of the theory and expression for the fermion density. The
polarization operator may be decomposed over three transversal tensors
\cite{Z95}:

        \be
        \Pi_{\mu\nu}(p) =
        \left(
        g_{\mu\nu} - \frac{p_\mu p_\nu}{p^2}           \right){\cal A}(p)
        +\left(
        \frac{p_\mu p_\nu}{p^2} -
        \frac{p_\mu u_\nu + u_\mu p_\nu}{(pu)} +
        \frac{u_\mu u_\nu}{(pu)^2} p^2                          \right)
        {\cal B}(p)
        + \imath \varepsilon_{\mu\nu\alpha}p^\alpha
        {\cal C}(p)~.
        \label{pimunu}
        \ee

The coefficients ${\cal A}$, ${\cal B}$ and ${\cal C}$ are the
functions of
$\mu, ~B, ~p_0$ and ${\bf p}^2$ only, $u=(1,0,0)$ is the 3-velocity of the
medium in the rest frame. The above coefficients have no divergencies except
for an infinite number of poles \cite{Z89} corresponding to the
electron-positron pair creation by a photon and fermion phototransition from
one Landau level to another.

With this expression for the polarization operator one may calculate the gauge
field propagator ${\cal D}_{\mu\nu}$ with corresponding corrections, ${\cal
D}_{\mu\nu}^{-1} = {D}_{\mu\nu}^{-1} - \Pi_{\mu\nu}$,

        \bea
        \lefteqn{{\cal D}_{\mu\nu}(p) =}\label{dmunu}\\
        &&
        \left\{
        \left(
        g_{\mu\nu} - \frac{p_\mu p_\nu}{p^2}           \right)
        \left(
        p^2 - {\cal A}(p) + (1 - \dfrac{p^2}{(pu)^2}) {\cal B}(p)   \right)
        +\left(
        \frac{p_\mu p_\nu}{p^2} -
        \frac{p_\mu u_\nu + u_\mu p_\nu}{(pu)} +
        \frac{u_\mu u_\nu}{(pu)^2} p^2                          \right)
        {\cal B}(p)                     \right. \nonumber \\
        &&
        \Bigg.
        + \imath \varepsilon_{\mu\nu\alpha}p^\alpha
        {\cal C}(p)                                     \Bigg\} \times
        \left((p^2 - {\cal A}(p))(p^2 - {\cal A}(p) +
        (1 - \dfrac{p^2}{(pu)^2} {\cal B}(p) ) - p^2  {\cal C}^2(p)
                                                        \right)^{-1}
        + \xi \dfrac{p_\mu p_\nu}{p^4} \nonumber
        \eea

\noi
(in MCS QED$_{2+1}$ one ought to change ${\cal C}$ to (${\cal C} -
\theta$)).

In the static limit $(p_0 = 0, {\bf p}^2 \rightarrow 0)$  the propagator
${\cal D}_{\mu\nu}$ may be written in $\xi = 0$ gauge as follows (cf.
\refc{Pisarski}):

        \bea
        \begin{array}{lr}
        {\cal D}_{00}(p_0 = 0, {\bf p}^2 \rightarrow 0) =
        - \dfrac{{\bf p}^2 + {\cal A}}
        {({\bf p}^2 + {\cal A})({\bf p}^2 + \Pi_{00}) + {\bf p}^2{\cal C}^2}
        &,
        \\
        & \\
        {\cal D}_{0i}(p_0 = 0, {\bf p}^2 \rightarrow 0) =
        - \i \ve_{ij}p_j
        \dfrac{{\cal C}}
        {({\bf p}^2 + {\cal A})({\bf p}^2 + \Pi_{00}) + {\bf p}^2{\cal C}^2}
        &,\\
        & \\
        {\cal D}_{ij}(p_0 = 0, {\bf p}^2 \rightarrow 0) =
        \left(\delta_{ij} -  \dfrac{p_i p_j}{{\bf p}^2} \right)
        \dfrac{{\bf p}^2 + \Pi_{00}}
        {({\bf p}^2 + {\cal A})({\bf p}^2 + \Pi_{00}) + {\bf p}^2{\cal C}^2}
        &,
        \end{array}
        \label{dstat}
        \eea

\noi
$\Pi_{00}= \dfrac{\bp^2}{p^2} {\cal A} + \dfrac{\bp^4}{p_0^2p^2}{\cal B}$. In
\Eq{dstat} ${\cal A}$, ${\cal C}$ and $\Pi_{00}$ are taken in the static limit.

Since ${\cal A}$ in the static limit is proportional to ${\bf p}^2$, one
actually needs to know two functions to describe the static limit of the gauge
field propagator, namely ${\cal C}$ and $\Pi_{00}$.  Fortunately, both these
coefficients may be obtained from the fermion density. First, it is well-known
\cite{F} that

        \be
         \Pi_{00} (p_0 = 0, {\bf p}^2 \rightarrow 0) = e^2
        \dfrac{\d \rho}{\d \mu}~~~.
        \ee

Then, it follows from the definition of the polarization operator
$\Pi_{\mu\nu}(x,x') = \i\frac{\delta <j_\mu (x)>}{\delta A_\nu (x')}$,
that in the static limit the components  $\Pi_{0j}$ ~($j=1,2$) are:

        \be
        \Pi_{0j}(p_0 = 0, {\bf p}^2 \rightarrow 0) =
        \Pi_{j0}^*(p_0 = 0, {\bf p}^2 \rightarrow 0) =
        - \imath  ~e \varepsilon_{ji}p_i\frac{\d
        \rho}{\d B}~~~.
        \ee

\noi
The latter yields  the induced Chern-Simons coefficient,

        \be
        {\cal C}(p_0 = 0, {\bf p}^2 \rightarrow 0) = e \dfrac{\d \rho}{\d
        B}\quad.
        \ee

Taking into account \Eq{rho} one has\footnote{We are omitting
$\sum\delta(\mu^2-m^2-2eBn)$ coming from
theta-functions derivatives since we are considering completely filled
Landau levels, $\mu^2\ne m^2+2eBn$.  In our forthcoming paper
a finite temperature case is studied, where $\rho$ and its derivatives
are smooth functions \cite{SZ}} $\Pi_{00}(p_0 = 0, {\bf p}^2 \rightarrow 0)=0$,
while   the Chern-Simons coefficient is  the following:

        \bea
        {\cal C}(p_0 = 0, {\bf p}^2 \rightarrow 0) = \frac{e^2}{4\pi} +
        \left\{
        \begin{array}{cl}
        {}~~\dfrac{e^2}{2\pi}
        \left[
        \dfrac{\mu^2 -m^2}{2eB} \right]
        ,\quad     & \mu>\/m;
        \\
        &\\
        0,                   \quad             & |\mu|<m;\\
        &\\
        {}-
        \dfrac{e^2}{2\pi}
        \left(
        \left[
        \dfrac{\mu^2 - m^2}{2eB}      \right]
        + 1      \right) ,\quad &\mu<- m.
        \end{array}\right.
        \label{thetaind}
        \eea

Since $\Pi_{00} (p_0 = 0, {\bf p}^2 \rightarrow 0)$ vanishes, in the
low-momentum limit the effective gauge field propagator of finite fermion
density Maxwell QED$_{2+1}$ with external magnetic field, \Eq{dstat} behaves
the same as the tree-level propagator of
MCS QED$_{2+1}$ with the Chern-Simons coefficient as in \Eq{thetaind}. Moreover,
the relation  $e\rho = \theta_{ind} B$ (effective equation of motion) holds true
with $\rho$ and $\theta_{ind}$ being as in Eqs. (\ref{rho}) and
(\ref{thetaind}), respectively.  Hence, in the low-momentum limit
finite-density magnetized Maxwell QED$_{2+1}$  is described effectively by
the MCS Lagrangian with massive gauge field with the Chern-Simons coefficient
proportional to the number of the filled Landau levels (at $B,T=0$ one has
${\cal C}= \frac{e^2}{4\pi}\theta(|m|-|\mu|)$, $\Pi_{00}= \frac{e^2}{2\pi}\mu
\theta (|\mu|-|m|)$).

\section{Summary}

We have demonstrated  that at zero temperature in Maxwell QED$_{2+1}$
magnetization is preferable at finite fermion density -- 2-dimensional
(parity-odd) fermion gas induces magnetic field.  Since the system with a
uniform magnetic field is described by the effective Maxwell-Chern-Simons
Lagrangian the magnetization  is rather  natural as fermions
coupled to Chern-Simons term become anyons -- particles carrying magnetic flux.

The possibility of  magnetization in the above-described system is intimately
connected to the fermion spectrum asymmetry in the external magnetic field: in
QED$_{2+1}$ with four-component fermions (and symmetric spectrum) the
magnetization is absent.  Since the asymmetry is of topological origin our
prediction based on the one-loop calculations will not be affected by higher
order corrections (the modified Coleman-Hill theorem \cite{LSW91,CH} which
claims that nonvanishing contribution to the Chern-Simons coefficient comes
from one loop only provides similar arguments). In this extent effects  in
a finite fermion density Maxwell QED$_{2+1}$ differs from those described in
MCS gauge theories \cite{H93,H95}, where  one-loop relations only fix
Chern-Simons coefficient, but relevant effects arise at two-loops level.

\section*{Acknowledgments}
I am grateful to A.E.Shabad for numerous discussions and valuable suggestions.
This work was supported in part by
RBRF grants $N^o$ 96-02-16210-a and 96-02-16117-a.


\end{document}